\documentclass[a4paper,10pt,twoside]{cpc-hepnp}
\usepackage{CJK,upgreek,fancyhdr}
\usepackage{multicol}
\usepackage{graphicx}
\usepackage{booktabs}
\usepackage{amssymb,bm,mathrsfs,bbm,amscd}
\usepackage[tbtags]{amsmath}
\usepackage{lastpage}
\usepackage{textcomp}
\usepackage{color}

\begin{document}
\begin{CJK}{UTF8}{gbsn}

\fancyhead[c]{\small To be published in Chinese Physics C}
\fancyfoot[C]{\small 010201-\thepage}

\title{Design and Development of JUNO Event Data Model\thanks{Supported by Joint Large-Scale Scientific Facility Funds of the NSFC and CAS (U1532258), the Program for New Century Excellent Talents in University (NCET-13-0342), the Shandong Natural Science Funds for Distinguished Young Scholar (JQ201402) and the Strategic Priority Research Program of the Chinese Academy of Sciences (Grant No. XDA10010900) }}

\author{%
      Teng Li$^{1;1)}$\email{liteng@hepg.sdu.edu.cn}%
\quad Xin Xia$^{1;2)}$\email{xiax@hepg.sdu.edu.cn}%
\quad Xing-Tao Huang$^{1}$\email{huangxt@sdu.edu.cn}%
\quad Jia-Heng Zou$^{2}$
\\
\quad Wei-Dong Li$^{2}$
\quad Tao lin$^{2}$
\quad Kun Zhang$^{2}$
\quad Zi-Yan Deng$^{2}$
}
\maketitle

\address{%
$^1$ School of Physics, Shandong University, Jinan, Shandong, 250100, China\\
$^2$ Institute of High Energy Physics, Chinese Academy of Sciences, Beijing 100049, China\\
}

\begin{abstract}
The Jiangmen Underground Neutrino Observatory (JUNO) detector is designed to determine the neutrino mass hierarchy and precisely measure oscillation parameters. The general purpose design also allows measurements of neutrinos from many terrestrial and non-terrestrial sources. The JUNO Event Data Model (EDM) plays a central role in the offline software system, it describes the event data entities through all processing stages for both simulated and collected data, and provides persistency via the input/output system. Also, the EDM is designed to enable flexible event handling such as event navigation, as well as the splitting of MC IBD signals and mixing of MC backgrounds. This paper describes the design, implementation and performance of JUNO EDM.
\end{abstract}



\begin{multicols}{2}

\section{Introduction}

The Jiangmen Underground Neutrino Observatory (JUNO) \cite{lab1, lab2} is a multi-purpose neutrino experiment being built in Guangdong, China. JUNO is designed to determine the neutrino mass hierarchy by precise measurement of the reactor antineutrino energy spectrum at a distance of 53 km from the Yangjiang and Taishan Nuclear Power Plants. The JUNO detector is capable of observing not only reactor neutrinos but also geoneutrinos, atmospheric neutrinos, solar neutrinos and neutrinos from supernova bursts.

\begin{center}
\includegraphics[width=6.5cm]{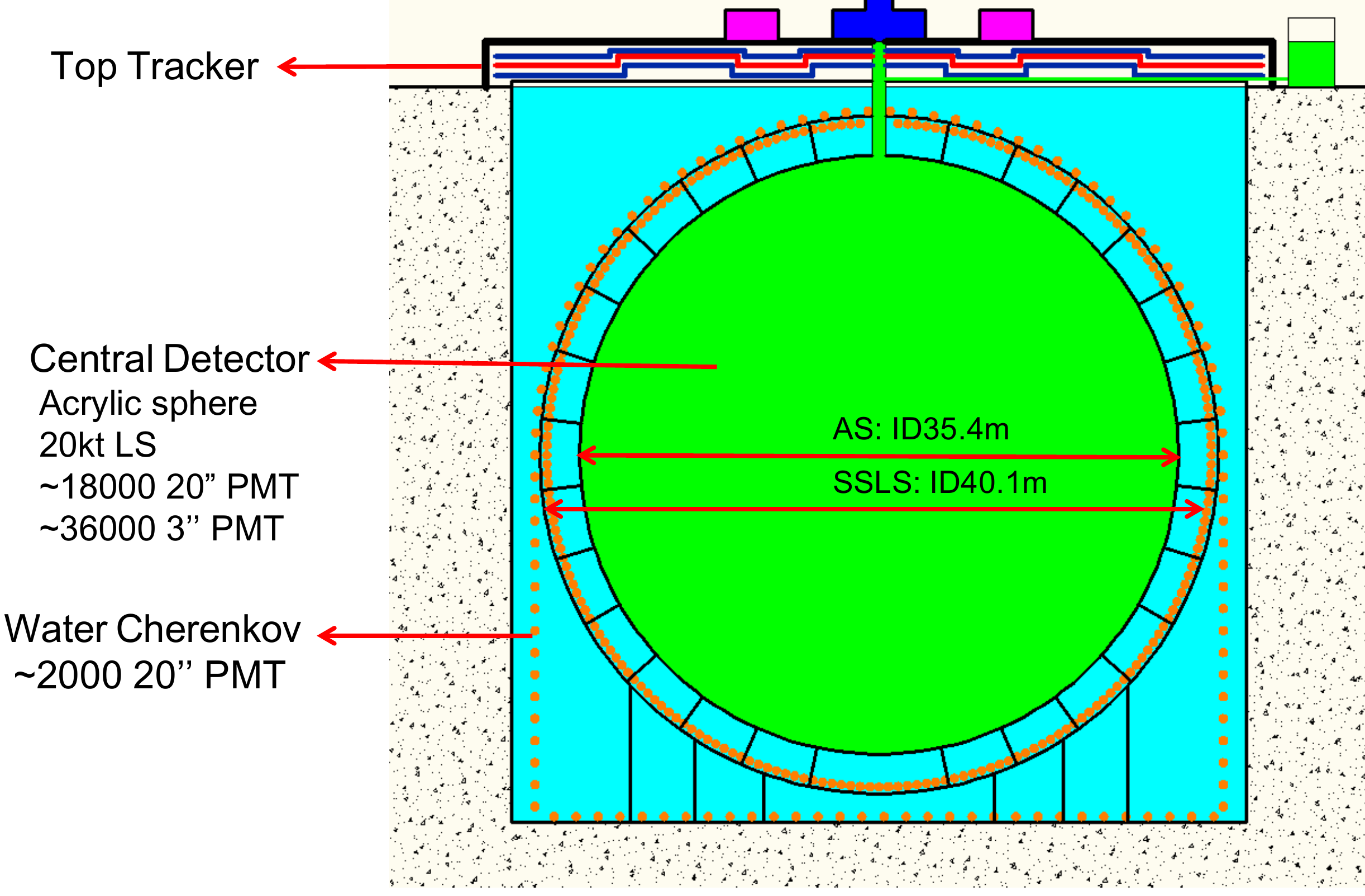}
\figcaption{\label{fig1}   Schematic view of JUNO detector. }
\end{center}

The JUNO detector is a liquid scintillator (LS) detector which consists of a central detector (CD) and a veto system, as shown in Fig. 1. The CD is an acrylic sphere filled with 20,000 tons LS and monitored with $\sim$18,000 20-inch photomultiplier tubes (PMTs) to reach an unprecedented energy resolution of 3\% at 1 MeV. The veto system consists of a Water Cherenkov detector and a Top Tracker system, which is used for muon detection, muon induced background study and reduction.

The JUNO offline software is designed to fulfill many requirements including Monte Carlo (MC) data production, raw data calibration, and event reconstruction as well as to provide tools for the physics analysis. The JUNO offline software is based on the general SNiPER (Software for Non-collider Physics ExpeRiment) framework \cite{lab3} with the main components implemented as SNiPER plugins, as shown in Fig. 2, and dependencies on external packages including ROOT \cite{lab4}, Geant4 \cite{lab5} etc.

\begin{center}
\includegraphics[width=8cm]{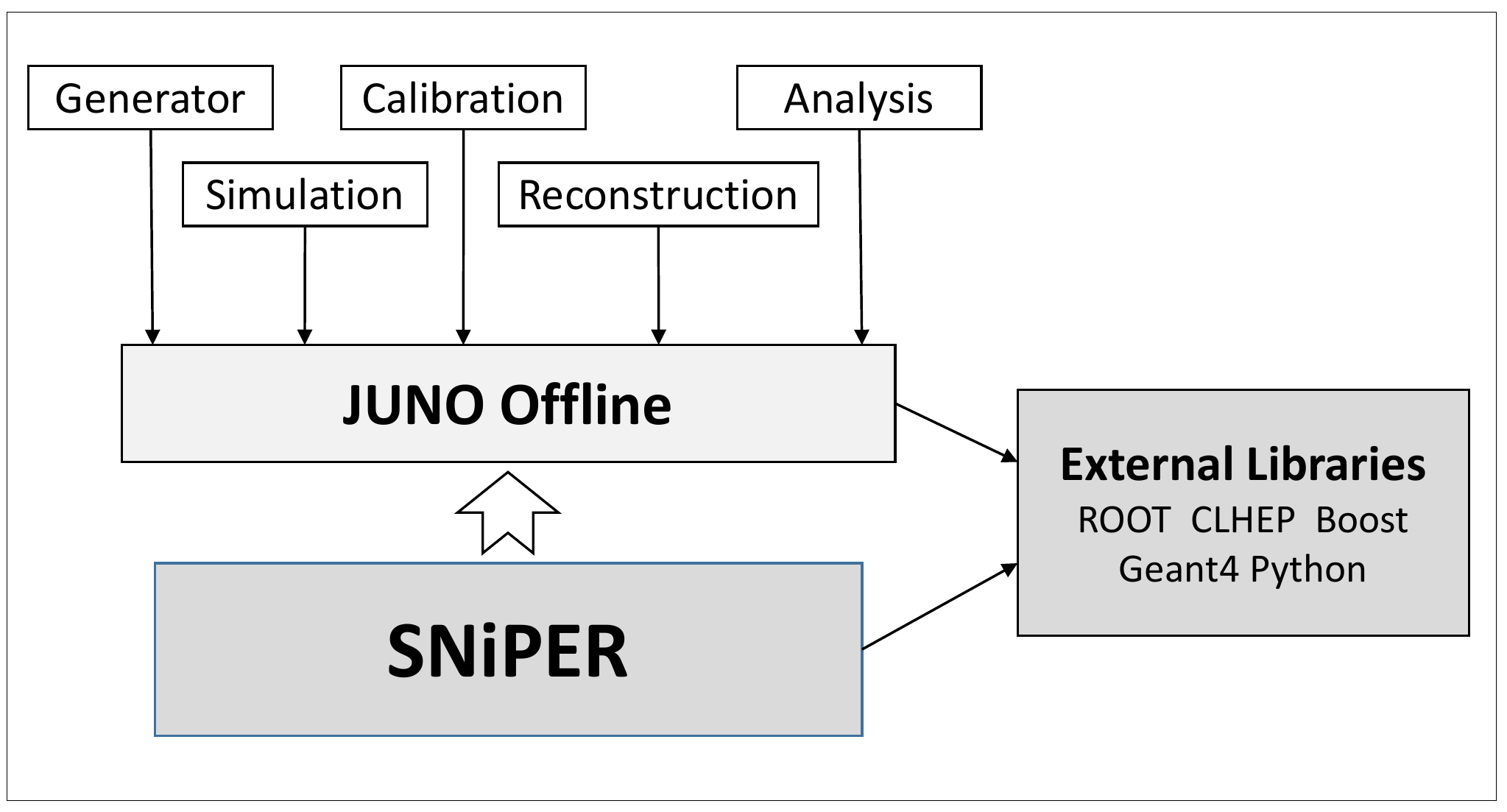}
\figcaption{\label{fig2}   Structure of JUNO offline software system. }
\end{center}

Fig. 3 illustrates the JUNO offline software workflow, with the data processing stages represented by diamonds and the event data they consume and produce represented by ovals. The event data is held in a dynamically allocated region of memory named DataBuffer. The event data at all stages can be persisted to files in the form of ROOT TTrees.

\begin{center}
\includegraphics[width=8cm]{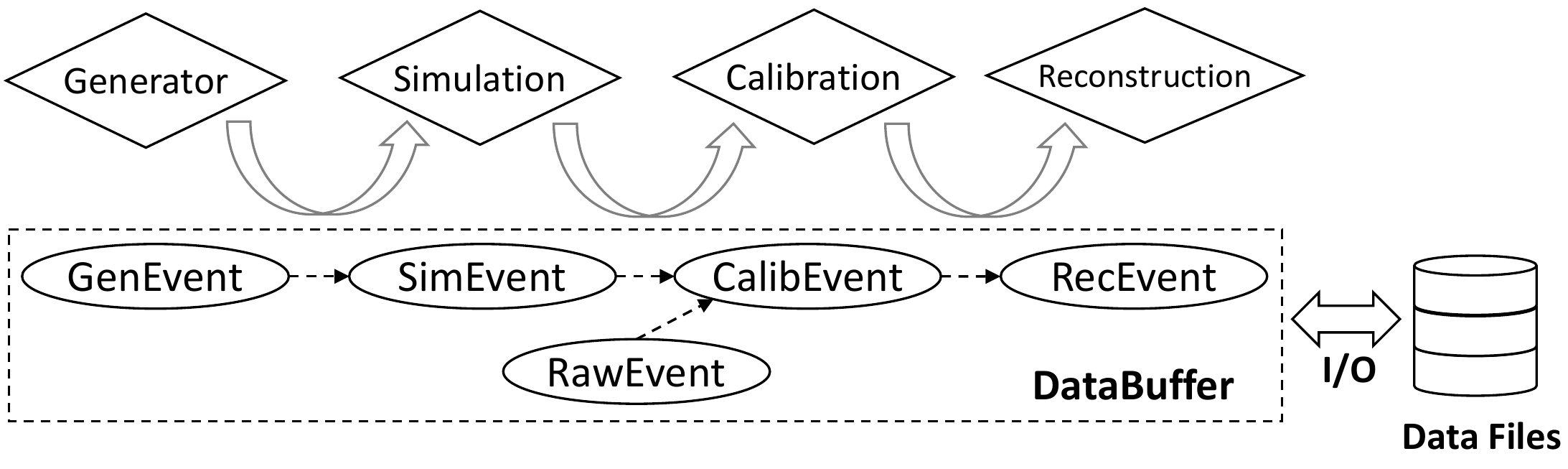}
\figcaption{\label{fig3}   Workflow of JUNO offline software system. }
\end{center}

For simulated data production, the Generator stage applies physics generators to produce GenEvent objects which are held in the DataBuffer. These GenEvent objects are used as the input to the Simulation stage which models the detector and electronics response. Similarly the Calibration and Reconstruction stages read data from the previous step and produce CalibEvent and RecEvent objects, respectively. All event data objects, such as GenEvent, SimEvent, RawEvent, CalibEvent and RecEvent, are defined and implemented by the JUNO Event Data Model (EDM). Also the ROOT based persisting data objects from all stages are defined and controlled by the EDM. Therefore, the EDM is a critical component for offline data processing and physics analysis.

\section{Requirements}

\subsection{Inverse Beta Decay reaction}

In JUNO, electron anti-neutrino signals are detected via the Inverse Beta Decay (IBD) reaction \cite{lab1}:

\begin{equation}
\overline{\nu}_{e}\,+\,p\,\rightarrow\,e^+\,+\,n
\end{equation}

\begin{center}
\includegraphics[width=6cm]{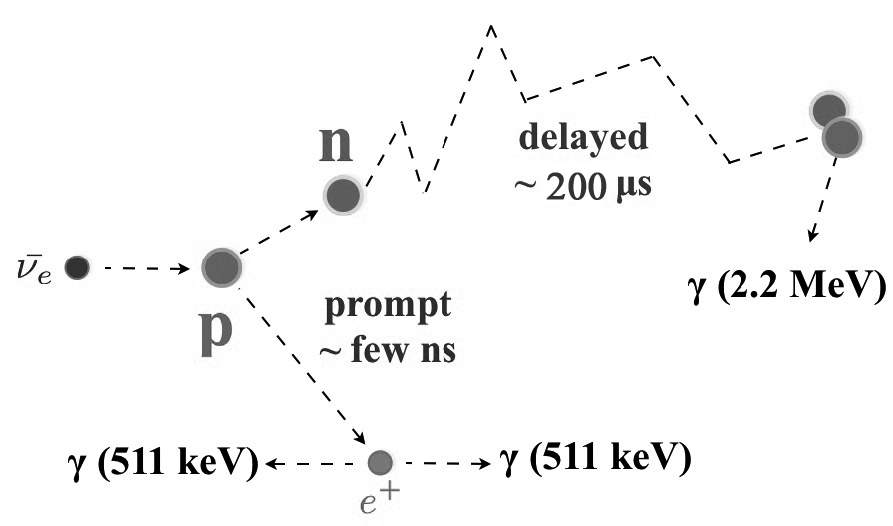}
\figcaption{\label{fig4}   Inverse Beta Decay reaction of the anti-electron neutrino. }
\end{center}

In the CD, an anti-electron neutrino interacts with a proton in the LS, producing a positron and a neutron. The positron promptly annihilates, producing two 511 keV gammas within a few ns. The neutron however is captured on free protons with a lifetime of $\sim$200 \textmu s, producing the delayed signal of a 2.2 MeV gamma. The time correlated prompt-delayed signals are triggered separately by the detector and are read out as separate events. Also, natural radioactive decays and cosmic muon induced processes yield backgrounds that potentially mix with or mimic the IBD signals.

\subsection{Specific requirements of JUNO}

The IBD reaction is the most important detection channel for the study of reactor neutrinos, supernova burst neutrinos, geoneutrinos, etc. Discrimination of the IBD signals from real data requires time coincidence, vertex correlation as well as an energy selection and a muon veto. However, for the MC data, a correlation mechanism of the simulated prompt-delayed signals can be used to make the process much easier.

When producing the MC simulation data, we firstly generate the IBD events and all kinds of backgrounds such as muon-induced backgrounds and radioactivity backgrounds separately at the detector simulation stage. Then at the electronics simulation stage, we split the IBD events and mix the backgrounds together according to their nominal event rates. This process requires a mechanism to correlate the prompt-delayed signals among the backgrounds to facilitate convenient analysis of the MC IBD event candidates. Moreover, several other detection channels \cite{lab2} in JUNO also lead to the prompt-delayed coincident signals, such as the charged-current interactions on $^{12}$C of supernova burst neutrinos and decay of protons. They will also benefit from such a correlation mechanism.

In order to study the detection efficiency and evaluate the systematic uncertainty during the physics analysis, the analysers may also need the event information of multiple processing stages (i.e. the reconstructed data and the MC truth of one event). Therefore, the EDM needs also to support the correlation between processing stages.

In addition, the large data volumes from $\sim$18000 PMTs necessitate the EDM to adopt efficient data access and storage techniques.

\section{Design}

\subsection{Design scheme}

JUNO EDM classes are implemented based on ROOT TObject in order to benefit from the features such as schema evolution, I/O streamers and runtime type identification. Also, such design is able to provide a familiar working environment for JUNO analysers, since ROOT has been very popular in high energy physics experiments.

Fig. 5 shows the design scheme of JUNO EDM. Separate metadata header object and event object pairs are used at each data processing stage in order to provide efficient data access tailored to the requirements of the users. The small metadata header objects derived from the HeaderObject contain only summary information such as tags, whereas the EventObject subclass instances contain the full infomation. Using header tags allows efficient event filtering by avoiding the loading of event data that is not required by users.

The association between header object instances and their paired event object instances is implemented via SmartRef \cite{lab6}, which provides control of the loading of the full data, as explained in section 4.2.

\end{multicols}
\begin{center}
\includegraphics[width=14cm]{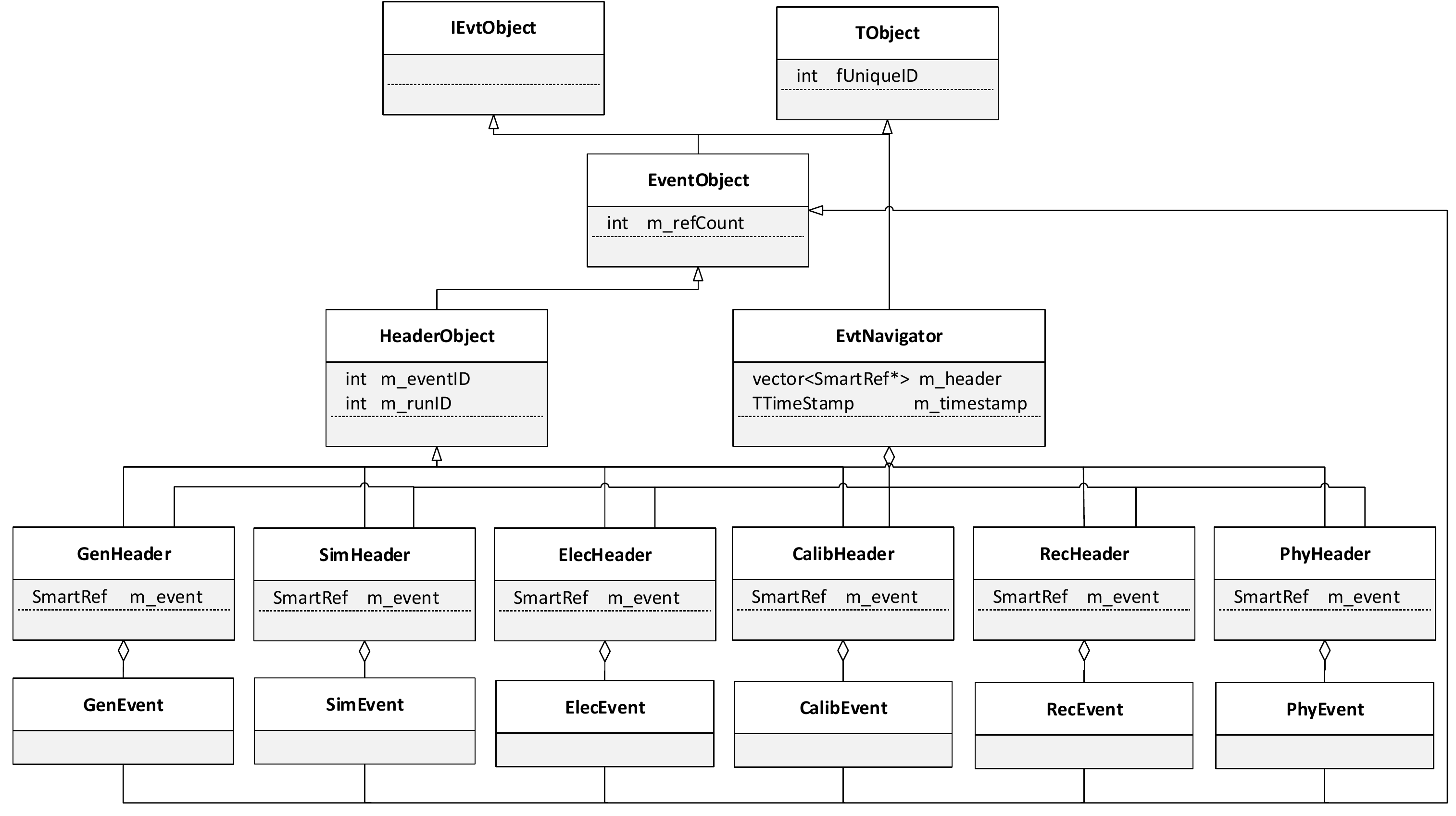}
\figcaption{\label{fig5} Design scheme of JUNO EDM.}
\end{center}
\begin{multicols}{2}

\subsection{Correlation mechanism}

Besides the association between the header and event objects, another two kinds of event data correlation are implemented via an EvtNavigator class, to facilitate correlated analysis between processing stages as well as the analysis of simulated IBD events. As shown in Fig. 6, an EvtNavigator instance keeps a list of SmartRef pointers that associate the event information from all the processing stages, such as Simulation, Reconstruction, etc. and allows itself to act as an index to the event data. For example, the EvtNavigator allows users to retrieve MC truth information corresponding to the reconstructed events. In addition to navigating within a single event, the EvtNavigator also allows navigation to prior and subsequent events.

\begin{center}
\includegraphics[width=8cm]{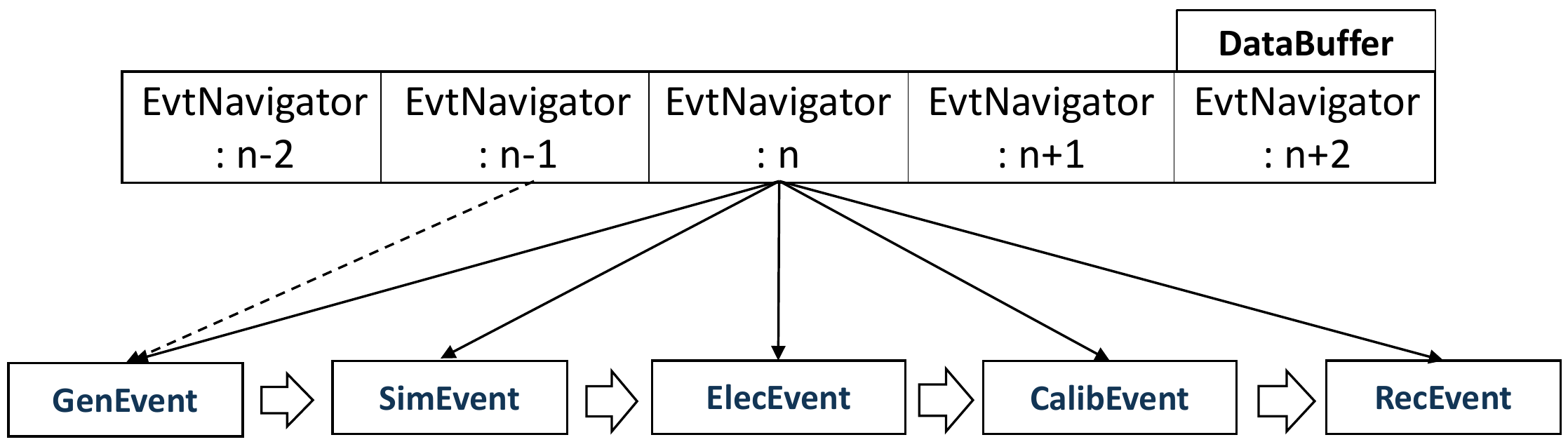}
\figcaption{\label{fig6} Schematic diagram of EvtNavigator.}
\end{center}

EvtNavigator is also capable of correlating the simulated IBD signals. Since the simulated IBD MC data is represented by single GenEvent and SimEvent objects at the generator and detector simulation stages, respectively. But at the electronics simulation stage, the simulated readouts split into separate the prompt-delayed ElecEvent objects. However, as the associations among multiple events and processing stages are retained by the EvtNavigator, and a path based interface is provided to follow the associations, users are able to trace the full processing history through the various stages and thus identifying the IBD signal pairs. As illustrated by the dashed line in Fig. 6, navigations from any processing stage to the corresponding GenEvent are implemented.

\section{Implementation}

The simple association between the header objects and the event objects as well as the correlation between processing stages both require flexible object referencing that survives even when the objects are persisted into multiple ROOT files.

\subsection{Smart reference}

\begin{center}
\includegraphics[width=3cm]{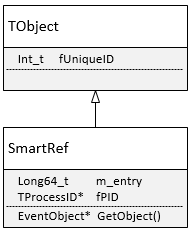}
\figcaption{\label{fig7}   The structure of the SmartRef. }
\end{center}

A smart pointer class named SmartRef is developed to provide persistable object referencing. The design of the SmartRef shown in Fig. 7 is based on the ROOT TRef \cite{lab7} but with some new features such as lazy loading and cross-file object referencing.

Every ROOT process creates at least one TProcessID instance which uniquely identifies the process in space and time by using a globally unique identifier TUUID \cite{lab8} instance. Also, the TProcessID instance maintains a pointer array which holds the EventObjects referenced by SmartRefs.

\begin{center}
\includegraphics[width=5.5cm]{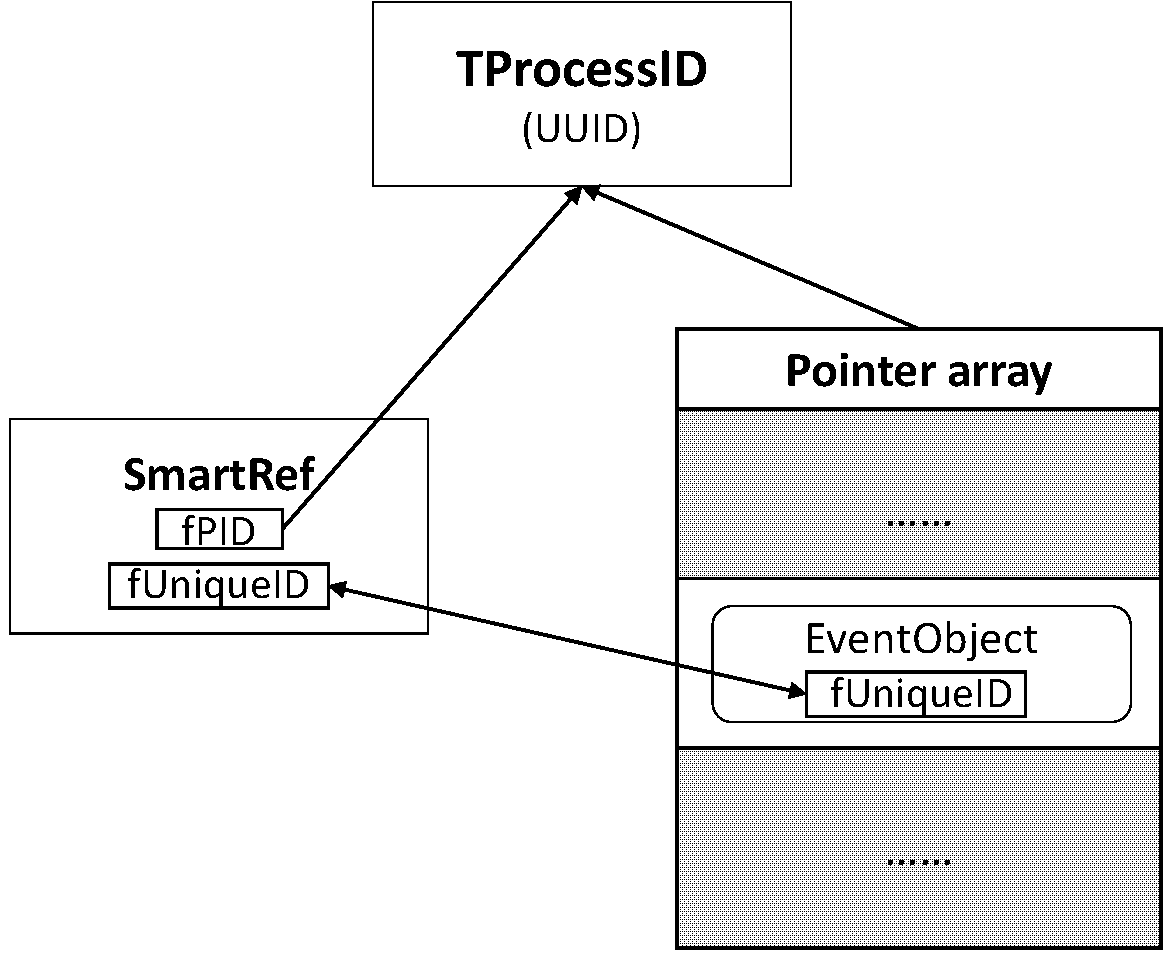}
\figcaption{\label{fig8}   Working principle of the SmartRef. }
\end{center}

When a SmartRef and a referenced EventObject are associated, a unique ID is assigned to both SmartRef and the referenced object, then the referenced object pointer is stored within the array maintained by the current TProcessID instance. For example, this happens when an EDM header object is associated to the paired event objects.

As shown in Fig. 8, the same unique ID held within the SmartRef and the referenced EventObject, and the uniqueness of the TUUID enable the association to be maintained both in memory and when being persisted to ROOT files. When a SmartRef or a referenced object is written out to a ROOT file, the corresponding TProcessID is also saved. When a ROOT file is opened in the input process, the TProcessID in the input file is loaded into memory, and the pointer array handled by this TProcessID is created again. After a referenced object is read in from the input file, it will be put into that pointer array accordingly to recover the association.

\subsection{Lazy loading mechanism}

The SmartRef supports a lazy loading feature compared to the ROOT TRef. Lazy loading means that the referenced object of SmartRef is not loaded from the input file until it is explicitly requested by the user, avoiding the performance overhead of data input until the data is actually needed.

The lazy loading mechanism is implemented within SmartRef, together with the ROOT I/O systems. During the output process, for each TTree holding the referenced EDM objects, a table named TablePerTree is created correspondingly. It includes the unique ID and branch ID of each referenced object saved in this TTree, which are used for reading the data back from ROOT files. The TablePerTrees are automatically built by the output service, and before the output file is closed, the current TProcessID instance and all TablePerTrees are written as metadata together with the data TTrees, as shown in Fig. 9.

\begin{center}
\includegraphics[width=8.5cm]{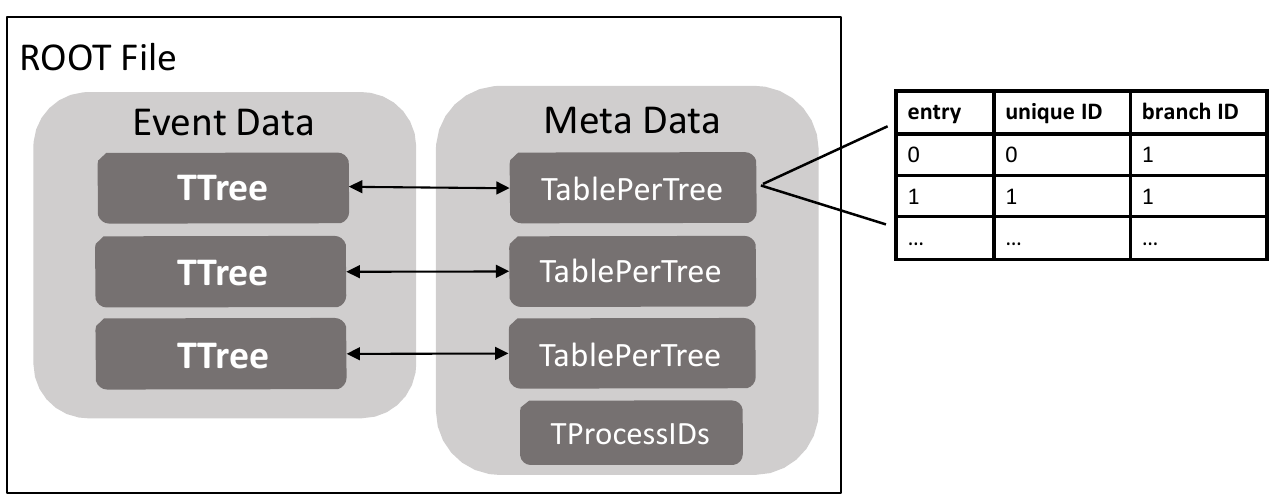}
\figcaption{\label{fig9} Schematic diagram of the ROOT file structure.}
\end{center}

During the initialization of the ROOT input system, a singleton module named InputElementKeeper that manages all input files and TTrees is created. It firstly scans all the input files quickly and reads their file metadata to construct a mapping between the TProcessID instances and the input files. When users attempt to access a referenced object via a SmartRef, it firstly searches the memory, and if the referenced object is not already loaded, the SmartRef queries the InputElementKeeper with its TProcessID to find the appropriate file. If the file is found, the SmartRef uses its unique ID to locate the position of the referenced object from the TablePerTrees, and loads it from this ROOT file.

\section{Performance}

Good performances including space and I/O efficiency are essential requirements for the offline software system. A series of tests have been performed to measure the performance. To avoid interference from unrelated processes the tests are executed on a single blade server (Intel(R) Xeon(R) CPU E5-2620 0 @ 2.00GHz) with all unnecessary processes suspended.

\subsection{Space consumption}

To quantify the size of the extra data for the lazy loading mechanism, a set of MC data samples is produced with the JUNO offline software containing from 2500 to 50000 events. The samples are saved in two formats: plain TTree and TTree with SmartRef and the required metadata. Fig. 10 shows the comparison of the file sizes of the two formats. With the same TFile compression level, SmartRef and required metadata add less that 1\% to the file size.

\begin{center}
\includegraphics[width=7cm]{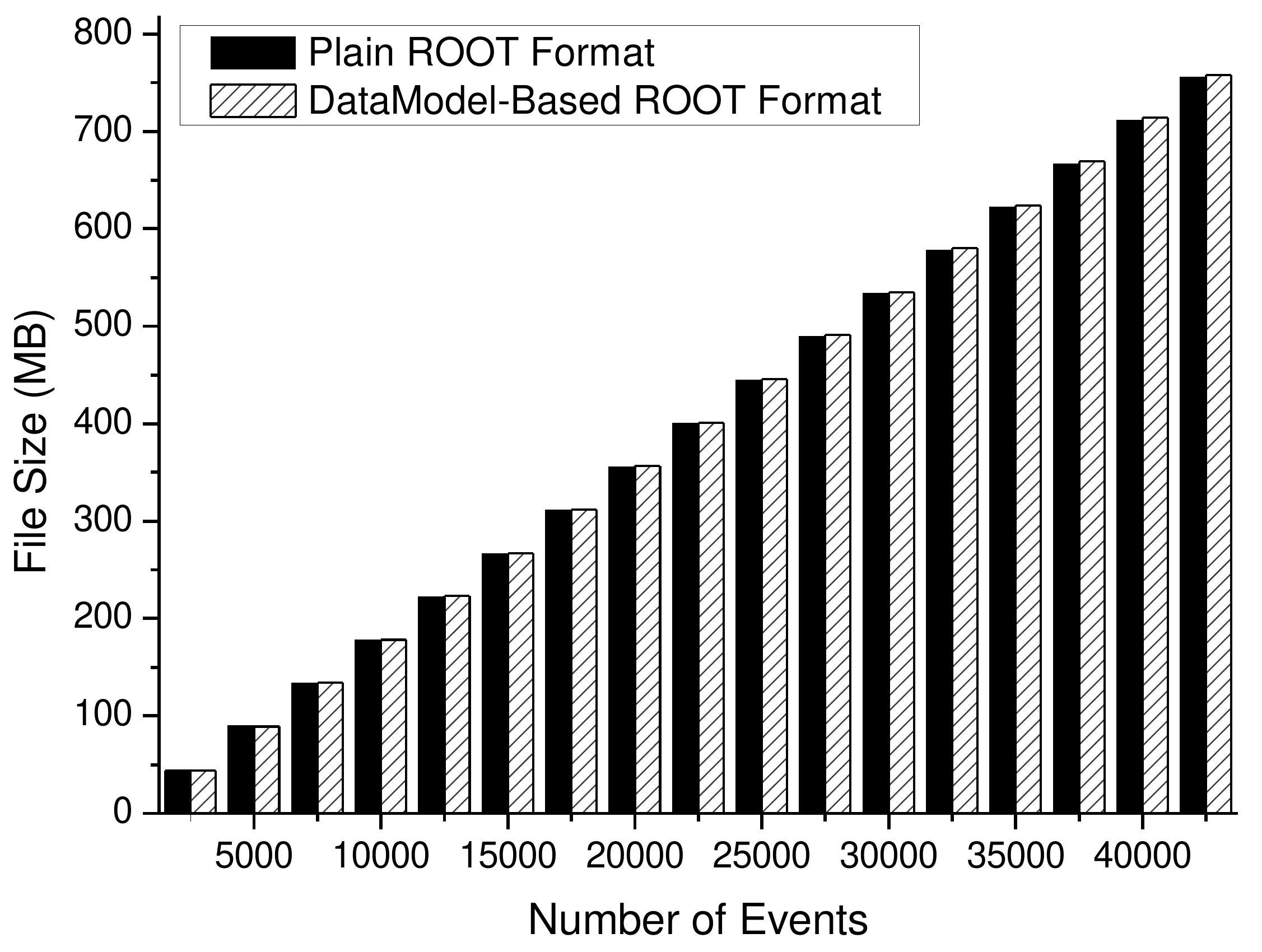}
\figcaption{\label{fig9} File size comparison of the plain TTree format and the persistent EDM format.}
\end{center}

\subsection{I/O efficiency}

The input performance with and without SmartRef lazy loading is measured by comparing file loading times for MC samples containing from 2500 to 50000 events. Fig. 11 shows the relative time difference between normal loading and lazy loading for four different selection fractions. To avoid statistical fluctuations an average of 100 measurements is used for each plotted point.

\begin{center}
\includegraphics[width=8cm]{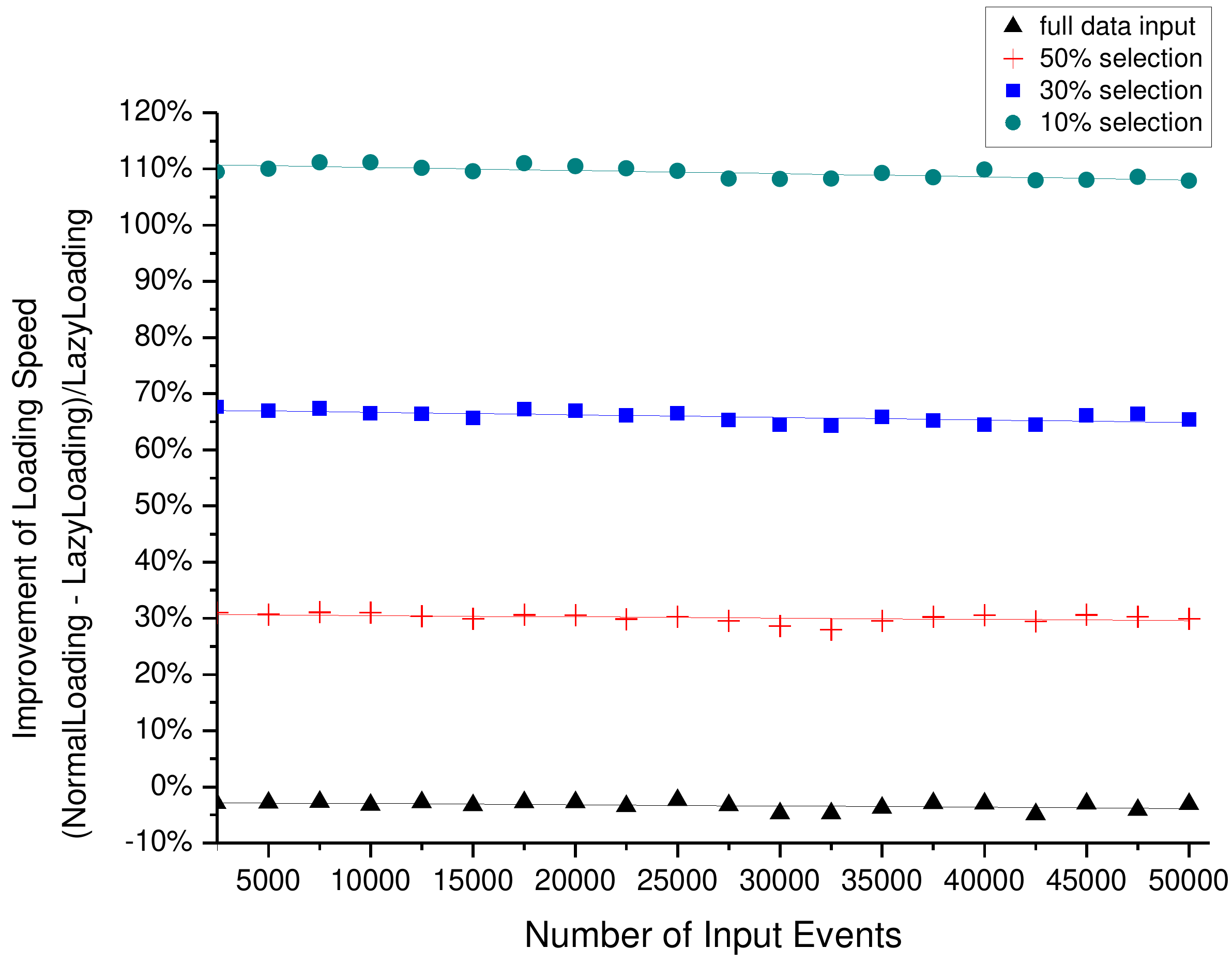}
\figcaption{\label{fig10} Comparison of data sample loading times.}
\end{center}

The loading times are found to scale linearly with the file size for all selections. The loading times without any selection are represented by the triangles in the plot, which show that the overheads caused by SmartRef and metadata are very small (less than 5\%). The crosses, squares and circles represent the relative difference of loading events with a random selection ratio of 50\%, 30\% and 10\% of the total respectively. The results show that the lazy loading mechanism provides an efficiency increase factor of about 30\%, 67\% and 110\%, respectively.

\subsection{Discussions}

The performance measurements show that lazy loading of SmartRef can provide a significant improvement on event loading time and cost only a minor increase of disk space for the necessary metadata. The improvement factor depends on the actual events accessed and the relative size of the Header and Event objects. JUNO analyses will typically require small selection fractions due to the high backgrounds compared to the neutrino signals, so lazy loading is foreseen to provide substantial performance improvements for most users.

\section{Conclusion and prospect}

The EDM introduced in this paper plays a central role in the JUNO offline software. It describes the event data entities through all processing stages for both simulated and collected data and provides persistency via the I/O system. To facilitate convenient analysis of simulated IBD events as well as the navigation between processing stages, the EvtNavigator and SmartRef classes have been developed. Based on the SmartRef, a lazy loading mechanism has been implemented to improve the performance of data I/O. The performance measurements show that the design of EDM and the lazy loading mechanism provide substantial performance improvements for event data loading.

Based on the JUNO EDM, the chain of the MC data production has been developed, including the detector and electronics simulation, the waveform reconstruction, the vertex and track reconstruction, etc.

The primary feature of this EDM design is the three types of event data correlation mechanisms based on the SmartRef. The correlation of prompt-delayed coincident signals is implemented for the special needs of neutrino experiments. However, the event navigation and association between paired header and event objects can be applied in both neutrino and other high energy physics experiments. The LHAASO and BESIII experiments have both applied our EDM design. For BESIII, we redesigned the EDM of reconstructed data based on the SmartRef, and through which the performance of the physics analysis has been greatly improved \cite{lab9}. It shows that such design pattern of EDM has great application prospects in high energy physics experiments, especially in future non-accelerator experiments, such as reactor neutrino, dark matter and double beta decay experiments.

\end{multicols}

\clearpage
\end{CJK}
\end{document}